\documentclass[12pt,a4paper,final]{article}
\usepackage[title,titletoc,toc]{appendix}
\usepackage{color}
\usepackage{hyperref} 
\usepackage[utf8]{inputenc}
\usepackage{amsmath}
\usepackage{amsfonts}
\usepackage{amssymb}
\usepackage{enumitem}

\usepackage{graphicx}
\graphicspath{ {images/} }
\usepackage[font=small,labelfont=bf]{caption}

\hypersetup{ colorlinks=true, linkcolor=blue, citecolor=red }
\newcommand{\hf}{{\frac{1}{2}}}
\newcommand{\p}{\partial}
\newcommand{\be}{\begin{equation}}                                             
\newcommand{\br}{\begin{eqnarray}}                                             
\newcommand{\ee}{\end{equation}}                                               
\newcommand{\er}{\end{eqnarray}}

\begin{document}

\title{
\hfill\parbox{4cm}{\normalsize IMSC/2017/03/02}\\
\vspace{2cm}
Exact Renormalization Group and Sine Gordon Theory}
\author{{Prafulla Oak\footnote {prafullao@imsc.res.in} } and B. Sathiapalan{\footnote{bala@imsc.res.in}} \\ \\
  \small Institute of Mathematical Sciences\\  
  \small CIT Campus, Taramani, Chennai-600113\\
  \small India }
\date{}
\maketitle
\begin{abstract}
The exact renormalization group is used to study the RG flow of quantities in field theories. The basic idea is to write an evolution operator for the flow and evaluate it in perturbation theory. This is easier than directly solving the differential equation. This is illustrated by reproducing known results in four dimensional $\phi^4$ field theory and the two dimensional Sine-Gordon theory. It is shown that the calculation of beta function is somewhat simplified. The technique is also used to calculate the c-function in two dimensional Sine-Gordon theory. This agrees with other prescriptions for calculating c-functions in the literature. If one extrapolates the connection between central charge of a CFT and entanglement entropy in two dimensions, to the c-function of the perturbed CFT, then one gets a value for the entanglement entropy in Sine-Gordon theory that is in exact agreement with earlier calculations (including one using  holography) in arXiv:1610.04233.
\end{abstract}
\newpage 
\tableofcontents 
\newpage

\section{Introduction}

The exact renormalisation group (ERG), first written down by Wilson \cite{WK,W,W2}  has been an object of much study. It has been developed further \cite{P} and different versions suitable for different purposes have been written down since then \cite{Morris,morris2,Wetterich}.  There are a large number of good reviews \cite{BB1,BB2,S1,IIS}. A lot of work has been done on the RG of the Sine-Gordon model over the last few years and many computations have been carried out analytically and numerically \cite{Yanagisawa:2016uia, Bacso:2015ixa, Kovacs:2014fwa, Wang:2014qka, Nandori:2013nda, Pelissetto:2012gv, Malard:2012iv, Nandori:2011ss, Nagy:2009pj, Nandori:2010xr, Nandori:2009ad, Nagy:2006se, Bozkaya:2005sc, Nandori:2003pk, Schehr:2003qc, Leclair:2003xj, Naon:2002qx, Kjaergaard:2000fh, Kehrein:2000ap, Park:1995fv, Ichinose:1994vf, Dimock:1991ig}.

This paper studies the application of ERG mainly to two dimensional field theories - the emphasis being on a simple way of writing down the solution to the ERG in terms of an evolution operator. Some known results are reproduced and a new result is obtained on the flow of the c-function in Sine-Gordon theory.
While the main application of the ERG has  been in the study of critical phenomena -  to obtain the numerical value of critical exponents,    our motivation comes from string theory.
In the context of string theory the renormalisation group has been used as a formal tool. Recently,  the ERG was used to obtain the equations of motion for the fields of the string somewhat as in string field theory\cite{BSERG}. In this approach string propagation in a general background is described as a completely general two dimensional field theory - all relevant, irrelevant and marginal terms are included. This is a natural generalization of the idea that a conformal field theory describes a consistent string background\footnote{See \cite{BSERG} for references to earlier papers on this topic.}. The condition of conformal invariance is imposed on the action. Thus the exact renormalisation group equation for this two dimensional theory is written down and the fixed point equations for the couplings are identified with the space time equations of motion of the background fields. One has to further generalise the original RG approach  to obtain equations that are gauge invariant. The new ingredient is the use of loop variables. In order to make the equations gauge invariant the two dimensional field theory is written in terms of loop variables \cite{BSLV}.  Loop variables have   also been incorporated into the ERG - and gauge invariant and interacting equations have been written down. Furthermore these equations are background independent \cite{BSERG}. 

Again within string theory, but now in the context of the AdS/CFT correspondence the idea of the renormalisation group has emerged in the guise of holographic RG \cite{Rangamani,Polchinski}. The RG flow of quantities has been equated with the evolution of the holographic dual bulk fields in the radial direction.
Thus a flow of renormalised coupling constants in the boundary is compared to the flow of the bulk field, which also requires renormalization. Many details of this comparison have been worked out in \cite{Morr2}. 
In field theory there are quantities such as the c-function of Zamolodchikov in two dimensions\cite{Zam} and the c and a-functions in four dimensions \cite{KS4d} that are monotonic along the flow. There have been attempts to find analogous quantities in the holographic dual in the bulk.
One such quantity is the entanglement entropy which has also been shown to be monotonic along the flow - both in the field theory and its holographic dual. (Although the precise connection with Zamolodchikov's c-function is not established.) 

Besides being interesting due to the connection with string theory, two dimensional models  have some advantages as an arena where these ideas can be developed. They are simpler to work with and their holographic dual $AdS_3$ equations are often exactly solvable. This motivates us to explore two dimensional field theories using ERG. 
An interesting and very non trivial field theory is the Sine-Gordon theory. The Sine-Gordon $\beta$-functions in fact are closely related to equations of motion of the bosonic string tachyon \cite{DS}. The equations of motion of the generalized version of the Sine-Gordon theory describes the bosonic string propagating in a tachyonic background and the $\beta$-function equations are proportional to the tachyon equation of motion. This is a special case of the connection to string theory mentioned above.
 
In the context of critical phenomena also this model has been related to a very interesting two dimensional model - the X-Y model. The X-Y model has an interesting phase transition first noticed by Kosterlitz and Thouless. It is possible to rewrite the X-Y model as a Sine-Gordon model. This theory has been studied in great detail in \cite{Amit} who obtained the phase diagram as well as the Kosterlitz-Thouless flow equations using continuum field theory techniques. They also showed that the model is renormalizable when physical quantities are written as a power series in terms of two coupling  constants\footnote{In the string theory context one of these couplings corresponds to the tachyon and the other to the dilaton.}.

In this paper we use the ERG to obtain the $\beta$ function equations of Sine-Gordon theory using the ERG. A particular form of the ERG due to Polchinski is used here. This ERG is reformulated as a linear evolution operator. Although this reformulation has been
noticed \cite{Morris,morris2}, it has not received much attention in practice. We show that it is very convenient to work out the flow of objects in a systematic perturbation series. In the usual continuum calculations $\beta$-functions are calculated as a byproduct of the renormalization program. As first explained by Wilson \cite{WK}, when one obtains the flow of a marginal coupling, in the limit that the UV cutoff is taken to infinity, the $\beta$-function has the property that it depends only on the value of the coupling and not explicitly on the scale. This also implies that the logarithmic divergence has the information about the $\beta$-function and higher orders in the logarithm are determined by the coefficient of the leading logarithmic divergence. Thus if we have an evolution equation one needs to only evaluate the leading divergence. Furthermore this is different from actually solving the ERG equation which gives a coupled differential equation involving an infinite number of couplings. The process of eliminating the irrelevant couplings and solving for the marginal coupling is automatically implemented during the perturbative evaluation of the evolution operator.

Thus mathematically one can imagine a set of coupled recursion equations \cite{WK} for a marginal coupling $g_l$, a relevant coupling $\mu_l$ and an irrelevant coupling $w_l$ obtained in a blocking transformation that implements the RG. We reproduce a summary of the discussion in \cite{WK}(The factors of 4 and 1/4 are illustrative):
\br 
g_{l+1} &=& g_l + N_g[g_l,\mu_l,w_l]\nonumber \\
\mu_{l+1} &=& 4 \mu_l + N_\mu [g_l,\mu_l,w_l] \nonumber \\
w_{l+1} &=& {1\over 4}w_l + N_w[g_l,\mu_l,w_l]
\er
Here $N_g,N_w,N_\mu$ are the nonlinear terms.
As explained in \cite{WK} one can reorganize the equations and solve them iteratively so that it depends on $w_0$ (initial condition for $w_l$) and $\mu_L$ (final value of $\mu_l$) and then one finds that on
solving this iteratively, and when $1<<l<<L$ is very large, so the memory of the initial conditions have been lost, one can set $\mu_L=w_0=0$
and obtain a recursion equation for $g_l$ alone
\be
g_{l+1}= V(g_l)
\ee

The crucial point is that in this limit $V(g_l)$ has no explicit dependence on $l$. One can now extract from this a $\beta$-function
$\beta_g= {dg\over dt}$ ($l$ is replaced by a continuous variable $t$) which depends only on $g(t)$. 

Now imagine using the evolution equation to obtain $g(t+\tau)$ starting from $g(t)$. One obtains a series of the form
\be 
g(t+\tau) = g(t) + \tau  {dg(t)\over dt} + {\tau^2\over 2!} {d^2 g(t)\over dt^2} +..
\ee
Now ${dg(t)\over dt}=\beta (g(t))$. Thus
\[
{d^2 g(t)\over dt^2}= {d\beta (g(t))\over dt}= {d\beta (g(t))\over dg}{dg(t)\over dt}={d\beta (g(t))\over dg}\beta(g(t))
\]
Thus when $t=0$, $\tau = \ln {\Lambda_0\over \Lambda}$ is what we call the logarithmic divergence in perturbation theory. What we are seeing is that the leading term decides the $\beta$-function and the higher powers of $\tau$ are fixed in terms of the leading term.\footnote{This is the counterpart of the statement for dimensional regularization, that the $1\over \epsilon$ pole determines the beta function, when only marginal couplings are present. The higher order pole residues are fixed in terms of the leading residue.} The application of the evolution operator in powers of the evolution Hamiltonian, gives us a series as above in $\tau$. It automatically gives the solution of the ERG recursion equations and one can extract a power series for the evolution of the marginal coupling. Thus $\beta$ -functions are obtained in a simple way without worrying about the technicalities of renormalization.

We illustrate this method with some examples such as the computation of the central charge of a free scalar field theory and calculating the flow of the coupling in $\phi^4$ theory in four dimensions. We then apply it to the more interesting case of the Sine-Gordon theory. We find that the equations obtained are consistent with those obtained in \cite{Amit}. While the precise coefficients are not the same, the combination of coefficients identified in \cite{Amit} as being universal, matches exactly. In addition to flow of couplings, one can study the flow of the c-function \cite{Zam,Cardy,Friedan, Komargodski}. In particular we do the calculation of the c-function for the Sine-Gordon theory.

Recently the entanglement entropy of this theory has been calculated both in the field theory and in the holographic dual and the answers are shown to agree to lowest order \cite{BBS}. The central charge calculation done here also gives results in exact agreement with these calculations - if we assume that the relation between entanglement entropy and central charge function persists at least to lowest non trivial order away from the fixed point. This computation is a first attempt towards developing an understanding  of a precise connection between the ERG in the boundary theory and Holographic RG in the bulk.

Another interesting computation, to understand better the holographic RG, would be to reproduce the $\beta$- functions for the Sine-Gordon model holographically. It has been shown in \cite{Sathiapalan:2017frk} that  an ERG equation  in a boundary theory  can be mapped to a
scalar field action in AdS space time. The main results are for a free theory. Some suggestions for how the interactions should work out were given there. To understand these issues better it is important to understand RG equations in the boundary theory and obtain them from some bulk computations. The precise connection between these equations and what is called ``holographic RG" - which is really a radial evolution equation of the bulk theory - needs to be understood better. These computations are a
step towards that goal. There is extensive literature on the AdS/CFT correspondence and holographic RG, \cite{Arutyunov:1998ve,Bourdier:2013axa,Maldacena:1997re,Kiritsis:2016kog} to name a few.

Once the perturbation is turned on it is no longer a CFT. This should reflect itself in the bulk deviations from AdS. This requires taking into account the gravitational back reaction. This back reaction in the bulk can be seen to manifest itself in the field strength renormalization of the boundary scalar fields. This gives us the beta function for the field strength renormalization. To compute this we look at the fluctuations of the graviton about the AdS. This contribution comes from another cubic vertex in the bulk. This is also equivalent to the dilaton equation in the string theory context.

In this paper we start by introducing the Polchinski equation for ERG and discuss the evolution operator at the gaussian point.  Then we show how the evolution operator takes the theory from the UV to the IR and integrates out irrelevant operators. Next, we demonstrate the exact solution of the ERG equation for the free theory. Then we compute the beta function of the $\phi^4$ theory in 4-dimensions. In the next section we compute the beta function for the Sine-Gordon theory. Next we discuss the central charge computations and compute the central charge for the Sine-Gordon theory and also show the computation of the change in the entanglement entropy as one flows to the gaussian fixed point. We end with summary and conclusions.

\section{Exact Renormalization Group}

\subsection{The Polchinski Equation}

Renormalization Group is integrating out high momentum modes leaving an effective theory of the low momentum modes. This is what is called "incomplete integration". Wilson observed that the equation($\dot{G}=\frac{\p G}{\p t}$)

\be \label{Wilson} 
{\partial \psi(X,t)\over \partial t} = -\hf \dot G{\p \over \p X}({\p\over \p X} + 2G^{-1}X)\psi(X,t)
\ee

realizes the notion of incomplete integration. The heat kernel of this equation gives a smooth interpolation of $\psi(X,t)$ between a completely unintegrated function $\psi(X,0)$ and its completely integrated form. Thus the equation is a possible candidate for an exact RG equation.

Substituting $\psi = e^{-S}$ in the above equation, we get

\be   \label{polch1}
{\p S\over \p t} = -\hf \dot G[{\p^2 S\over \p X^2} - ({\p S\over \p X})^2 + 2 G^{-1} X {\p S\over \p X}] + \underbrace {\dot G G^{-1}}_{fld~indep}
\ee

From here on we drop all field independent terms as they contain no dynamics and are vacuum bubles. Adding such terms shifts the energy level of the Lagrangian, like a cosmological constant, and does not change anything till you couple it to gravity.

If we substitute $\psi = e^{-\hf G^{-1}X^2}\psi'$ we get an equation: 
\be	\label{Polch}
 {\p \psi'\over \p t} = -\hf \dot G {\p^2 \psi'\over \p X^2}
\ee
Here $\psi' = e^{-S_{int}}$, $S_{int}$ is the interaction part of the action. Again in terms of $S_{int}$ it becomes an equation in the form first written by Polchinski \cite{Polchinski}
\be \label{polch2}
{\p S\over \p t}=-\hf \dot G[{\p^2 S\over \p X^2} - ({\p S\over \p X})^2]
\ee

In these equations one can replace $X$ by $\phi(p)$ and easily generalize to field theory. In a field theory RG $t$ is the logarithm of the ratio of scales:  the short distance cutoff $a(0)$ is changed to $a(0)e^t$. In a field theory action $\hf G^{-1}X^2$ would stand for the kinetic term (and $G$ for the Green function) and then $S$ would be the interaction part of the action.  Polchinski's equation is usually used in the form \eqref{polch2} (or in the form \eqref{polch1} for the full action). 

In this paper however we use it in the form \eqref{Polch}. This is a {\em linear} equation and is just a free particle Schroedinger equation. The formal solution of this equation in terms of an evolution operator  can easily be written down. 
Writing a formal solution in this form is useful in some situations: The ERG as is usually written down is an infinite number of equations that give the $\beta$-function of one coupling parameter in terms of  all the other infinite number of coupling parameters. The usual continuum beta function involves only a few of the parameters involving the lower dimensional operators. To go from the first form to the second form one has to solve these infinite number of equations iteratively \cite{WK}. The evolution operator method does this operation in a convenient way (as will be shown). It thus acts as a bridge between the ERG and the continuum field theoretic $\beta$-function.  

 \subsection{Free Theory}
 
 Let us understand the connection between the ERG equation and the evolution operator by considering the free theory
 as a pedagogical exercise. The first step is to construct the field theoretic version of Polchinski's ERG:
 
 \subsubsection{ERG and $\beta$-function}

The ERG acting on $\Psi$ is:
\be    
{\p \Psi\over \p t}= -\hf\int dz~ \int dz'~ \dot G(z,z',t) {\delta^2\Psi\over \delta X(z) \delta X(z')} \equiv -H(t) \Psi(t)
\ee 
 with $\Psi = e^{-\int du~L(u,t)}$. This can be written as an ERG for $L$. 
 
 Let us write the Wilson interaction as $S \equiv -\int du~L(u,t)$. 
 We get
 \be  \label{ERG1}
 {\p S \over \p t}= -\hf\int dz_1~ \int dz_2~ \dot G(z_1,z_2,t) [{\delta^2 S \over \delta X(z_1) \delta X(z_2)} + {\delta S\over \delta X(z_1)}{\delta S\over \delta X(z_2)}] 
 \ee
 
We could start with a local bare action:
 \[
S= -\int du~ \hf \delta m^2(u) X(u)^2
 \]
where  $\delta m^2(u)= (e^{2\phi(u)}-1)m^2$  is a position dependent coupling(mass), but in general even if we start with a local action, after one iteration of the RG it becomes non-local. So we start with a non-local action
\be
S = -\int du \int dv ~\hf z(u,v,t) X(u) X(v) - m_0(t)
\ee

Substituting this in \eqref{ERG1} we get

\br   \label{ERG2}
\dot m_0(t) &=& -\hf \int _{z_1}\int _{z_2} \dot G(z_1,z_2,t) z(z_1,z_2,t) \nonumber \\ 
\dot z(u,v,t) &=&  \int _{z_1}\int _{z_2} \dot G(z_1,z_2,t)z(z_1,u,t)z(z_2,v,t)
\er

The set of $\beta$-function equations \eqref{ERG2} is exact.   But the simplicity is a little misleading because $z(u,v,t)$ is a function of two locations $u,v$ and actually represents an infinite number of  local (position dependent) coupling functions, which can be defined by Taylor expansions. Note that even for the free field case we get a non local Wilson action.

 \subsubsection{Evolution Operator }

\eqref{Polch} can be written as 
\be
{\p \psi\over \p t} = -H\psi
\ee
with $H = \hf\dot G {\p^2\over \p X^2}$, 
for which the solution is formally

\be\label{formal}
\psi(X,t)=e^{-\int _0^tdt' H}\psi(X,0)=e^{-\hf (G(t)-G(0)) {\p^2\over \p X^2}}\psi(X,0)
=e^{-\hf (F(t)) {\p^2\over \p X^2}}\psi(X,0)
\ee

Consider the Schroedinger equation:
\[
{\p \psi\over \p T} = -\hf F(t){\p^2\over \p X^2}\psi
\]
which is solved formally as

\[
\frac{d\psi}{\psi}=-\hf  \bigg(\int dT \bigg) F(t) \frac{\partial^2}{\partial X^2}  
\]

\[
\log \psi(X,t,T)=-\hf T F(t) \frac{\partial^2}{\partial X^2}   + \log \psi(X,0)
\]

\[
\psi(X,t,T)=e^{-T \hf F(t){\p^2\over \p X^2}}\psi(X,0)
\]
With $T=1$ we get our solution \eqref{formal}.
The solution to the schrodinger equation is known in terms of a kernel
\[
\psi (X,t,T)=\int dX'~e^{ {1\over 2F(t)}{(X-X')^2\over T}}\psi (X',0)
\]
So setting $T=1$ we get the solution to our original problem:
\be\label{ergintegral form}
\psi (X,t)=\int dX'~e^{ {1\over 2F(t)}(X-X')^2}\psi (X',0)
\ee
If we write $\psi=e^{-S}$ we get
\be
e^{-S(X,t)}=\int dX'~e^{ {1\over 2F(t)}(X-X')^2}e^{-S(X',0)}
\ee
which can also be written in a well known standard form as  \cite{Sonoda,Morris,morris2,IIS}
\be
e^{-S(X,t)}=\int dX'~e^{ {1\over 2F(t)}X'^2}e^{-S(X+X',0)}
\ee

 We can convert the above solution to a field theoretic case and in the free theory, obtain an  exact form of the solution to ERG evolution. Working in momentum space, all we need to do is to replace $X$ by $X(p)$. The integral over $X'$ becomes a functional integral over $X(p)$ and in the action we need to sum over all $p$. The ''propagator'' $F(t)$ becomes $F(p,t) = G(p,a(0)e^t)-G(p,a(0))$:
\be
\int {\cal D}X' e^{\hf \int {d^2p\over (2\pi)^2}F^{-1}(p)X'(p)X'(-p)} e^{-S[X+X']}
\ee

In this form it looks a free particle (field) calculation where the propagator is $F(p,t)\equiv G(p,a(t))-G(p,a(0))$ with $a(t)=a(0)e^t$ the moving cutoff. Thus the propagator
only propagates the modes that are being integrated out. So, for e.g., when $t=0$ it vanishes because no integration has been done.

\subsubsection{Free Field Theory: Exact Solution of ERG}

In the case of the free field the integrations can be carried out exactly.

\begin{equation}
\int {\cal D}X' e^{\hf \int {d^2p\over (2\pi)^2}F^{-1}(p)X'(p)X'(-p)} e^{-\hf \int z(p) (X+X')(p)(X+X')(-p)}
\end{equation}

\begin{equation}\label{befint}
=\int {\cal D}X' e^{-\hf \int {d^2p\over (2\pi)^2}\underbrace{(z(p)-F^{-1}(p))}_{{\cal F} ^{-1}(p)}X'(p)X'(-p)} e^{-\hf \int_p z(p) X(p)X(-p) + 2 z(p) X(p) X'(-p)}
\end{equation}

\begin{equation}
=Det^{\hf}[{\cal F}] \exp[-\hf \int_p X(p) \underbrace{({z\over 1-Fz})}_{-z(t)}X(-p)]
\end{equation}

Here

\[
z(t)=-{z\over 1-Fz}
\]

and

\[{\cal F}=\frac{F}{Fz-1}
\]

Thus we have an exact solution for the Wilson action. 

Furthermore,
\[
{dz\over dt} = z^2 {dF\over dt}=z^2 {dG\over dt}
\]
which is the second eqn in \eqref{ERG2}. We thus make contact with the differential version of ERG.

\subsection{$\beta$-function of $\phi^4$ theory in four dimensions}
Now we illustrate the method of calculating the $\beta$ function for the $\phi^4$ theory using the ERG.
We use the Polchinski equation with all the corrections to kinetic term being put into the interactions. Since we are only integrating modes with  $p>\Lambda$ we do not need a mass as a regulator. So we can put $m^2=0$.

The evolution operator is
\[
e^{-\hf \int _{x_1} \int _{x_2} (G(x_1,x_2,\Lambda(t))-G(x_1,x_2,\Lambda(0))){\delta^2\over \delta \phi(x_1)\delta \phi(x_2)}}
\]

We set 
\[
\psi(0) = e^{-S[\phi,0]}=e^{-{\lambda\over 4!} \underbrace{\int _x \phi(x)^4}_{''V''}}
\]

The action of the evolution operator on $e^V$ is,
\[
e^{-\hf \int _{x_1} \int _{x_2} F(x_1,x_2){\delta^2\over \delta \phi(x_1)\delta \phi(x_2)}}e^{-{\lambda\over 4!}\int \phi^4}
\]

We can keep some terms in the exponent and bring down the rest:
\[=
\int {\cal D}\phi' e^{\hf \int _{x_1}\int_{x_2}F^{-1}(x_1,x_2)\phi'(x_1)\phi(x_2')}e^{-{\lambda\over 4!}\int (\phi^4 + 4 \phi^3 \phi' + 6\phi^2\phi'^2)}\]\[[1-{\lambda\over 4!}\int( 4 \phi \phi'^3 +\phi'^4) +{1\over 2!} ({\lambda\over 4!})^2[\int ( 4 \phi \phi'^3 +\phi'^4)]^2+...]
\]

Let us evaluate:
\[
\int {\cal D}\phi' e^{ \int _{x_1}\int_{x_2}\underbrace{(\hf F^{-1}(x_1,x_2)-{\lambda\over 4!}  \delta (x_1-x_2) 6 \phi^2(x_1))}_{\hf H^{-1}(x_1,x_2)}\phi'(x_1)\phi(x_2')-\int_x  J(x) \phi'(x)}
 \]

\[=
Det^{-\hf}H^{-1}(x_1,x_2) e^{\hf \int _{x_1} \int_{ x_2} J(x_1)H(x_1,x_2)J(x_2) }
\]

\[
Det ^\hf H = e^{\hf Tr\ln [{1\over F^{-1}-{2\lambda \over 4!} 6 \phi^2 I}]}
\]
\[=
e^{\hf Tr \ln F - \hf Tr\ln [1-{\lambda \over 2}\phi^2 F]}
\]

Expand the log:
\[
\hf Tr \ln [1-{\lambda \over 2}\phi^2 F]=\]\[\hf \bigg( -{\lambda \over 2} \int _x \phi^2(x) F(x,x) - \hf ({\lambda \over 2})^2 \int _{x_1}\int _{x_2} \phi^2(x_1) F(x_1,x_2)\phi^2(x_2)F(x_2,x_1) +...\bigg )
\]

In momentum space $F$ can be understood as a propagator with momentum restricted in the range $\Lambda<p<\Lambda_0$.
Thus
\[
\phi^2(x)F(x,x)= \phi^2(x)\int {d^4p\over (2\pi)^4} {1\over p^2}
\]
This is the usual quadratically divergent mass correction.
To get the correction to the $\phi^4$ term we consider the next term in $-\hf Tr \ln [1+{\lambda \over 2}\phi^2 F]$,
\[
-\hf \times \hf {\lambda^2\over 4} \times {1\over (4\pi)^2} \int _{\Lambda_0^2}^{\Lambda^2} p^2 dp^2 {1\over p^4}\phi(0)^4
\]
The external momentum is set to zero i.e. $\phi(x)$ is uniform. This is a correction to $\lambda\over 4!$ so we factor out $4!$ to get,
\[-
{4!\over 4!}\times\hf \times \hf {\lambda^2\over 4} \times {1\over (4\pi)^2} \int _{\Lambda_0^2}^{\Lambda^2} dp^2 {1\over p^2}
\]
\[
=-
{1\over 4!}{3\over 2} {\lambda_0^2\over (4\pi)^2}\ln {\Lambda^2\over \Lambda_0^2}
\]
 Since $\Lambda = e^{-t}\Lambda_0$ we get
 \[-
 {1\over 4!}{3\over 2} {\lambda^2\over (4\pi)^2} (-2t)
 \]
 Thus

 \[
- \lambda(t)= -{\lambda\over 4!} + {3\over (4\pi)^2}\lambda^2t
 \]
 \[
 \dot \lambda =-{3\over 16\pi^2}\lambda^2
 \]

This is the well known $\beta$ function of the $\phi^4$ theory in four dimensions. What about contributions to $\beta$ function from $\langle {\lambda \over 3!}\phi \phi'^3 \rangle$? For this we calculate,
\[
\int {\cal D} \phi' [-{\lambda \over 3!}\phi \phi'^3] e^{ \int _{x_1}\int_{x_2}{\hf H^{-1}(x_1,x_2)}\phi'(x_1)\phi(x_2')-\int_x  J(x) \phi'(x)}
\]
where $J$ will be set to ${\lambda \over 3!}\phi^3$ in the end. Thus one evaluates
\[
-{\lambda \over 3!}\phi {\delta^3\over \delta J(x)^3}[Det^{-\hf}H^{-1}(x_1,x_2) e^{\hf \int _{x_1} \int_{ x_2} J(x_1)H(x_1,x_2)J(x_2) }]
\]
All terms necessarily have one factor of the form $HJ$. To lowest order in $\lambda$, $H=F$. When we set $J={\lambda \over 3!}\phi^3$ the external momentum is zero (for constant $\phi$) and thus we have an $F$ propagator with zero momentum. This is zero because $F$ is non zero only for   momenta greater than $\Lambda$. Thus this correction is zero to lowest order.

\section{The Sine-Gordon theory.}

We now turn to the Sine-Gordon theory. We compute the $\beta$-functions for this theory using the ERG evolution operator.

The action for the theory is

\begin{equation}
S= \frac{1}{4 \pi}  \int \frac{d^2x}{a(0)^2}  \bigg((\partial X)^2+m^2X^2+F \cos(bX)  \bigg)
\end{equation}

$a(0)$ is the UV cut-off.

\subsection{The Green's Function.}

The Green function for the Klein Gordon field in two dimensions in Euclidean space is
\be  \label{GF}
G(x_2,t_2;x_1,t_1)=\int _0^\infty ds~ (\frac{1}{{4\pi s}}) e^{- m^2 s} e^{-\frac{(x_2-x_1)^2+(t_2-t_1)^2}{4s}}
\ee
The small $t$ region gets contribution from $x^2=0$ region. This is the UV. A way to regularise this is to cutoff the integral:
\be	\label{RegGF}
G(x_2,x_1,\epsilon)=\int _\epsilon^\infty ds~ (\frac{1}{{4\pi s}}) e^{- m^2 s} e^{-\frac{(x_2-x_1)^2}{4s}}
\ee
\be	\label{GF1}
\int _0^\infty ds~ (\frac{1}{{4\pi s}}) e^{- m^2 s} e^{-\frac{(x_2-x_1)^2+(t_2-t_1)^2}{4s}}= {1\over 2\pi} K_0(mx)
\ee

Here $x=\sqrt{(x_2-x_1)^2+(t_2-t_1)^2}$.

\br
&=& {1\over 2\pi}\sqrt{{\pi\over 2 mx}}e^{-mx}+....: mx>>1\nonumber \\
&=&{1\over 2\pi}[-ln~(mx/2)(1+ \sum _{k=1}^\infty {(mx/2)^{2k}\over (k!)^2})]+ \psi (1) + \sum _{k=1}^\infty {(mx)^{2k}\over 2^{2k}(k!)^2}\psi(k+1)
~~~: mx<<1\nonumber
\er

We will do our calculations in the $mx<<1$ regime.

\subsection{ Reproducing the continuum $\beta$-functions.}

We would like to reproduce the flow for F and $b$. For the kinetic term 
\[
S_{Kinetic}= {1\over \alpha'}\int d^2z ~\partial_z X \partial _{\bar z} X = {1\over 2\alpha'}\int d^2x ~\partial_a X \partial^aX
\]
the Green's function in complex coordinates is 
\[
G=< X(z) X(w)> = -{\alpha'\over 2\pi} \ln {|z-w|+a(0)\over R}
\]

and we choose $\alpha'=2\pi$ and substitute that when we carry out calculations in the later sections. Here R is some scale.

\[
<X(0) X(0)>=-{\alpha'\over 2\pi} \ln~{a(0)\over R}
\]
In real coordinates

\begin{equation}
G=< X(x_1) X(x_2)> = -{\alpha'\over 4\pi} \ln |{(x_1-x_2)^2+a(0)^2\over R^2}|
\end{equation}

The evolution operator acting on the unintegrated theory gives
\br
\psi(t) &=&e^{-\int _{t_0}^tH(t')dt'}\psi(0) \nonumber\\
&=& e^{-\hf \int _{t_0}^tdt'\int d^2x_1~\int d^2x_2~\dot G(x_1,x_2,t'){\delta^2\over \delta X(x_1)\delta X(x_2)}}e^{-S[X,0]}\nonumber\\
&=& \int {\cal D}X''e^{\hf \int d^2x_1~\int d^2x_2~F^{-1}(x_1,x_2)X''(x_1)X''(x_2)}e^{-S[X+X'']}
\er

Here
\[
S[X,0]= \int {d^2x\over a(0)^2} \frac{F}{4 \pi} [{e^{ib X}+ e^{-ib X}\over 2}]\]

and

\[
F(x_1,x_2,t)= G(x_1,x_2,a(t))-G(x_1,x_2,a(0)) =G(x_1,x_2,a(0)e^{t})-G(x_1,x_2,a(0) )
\]

Thus
\begin{equation}
\label{fuvt}
F(x_1,x_2,t)= -{\alpha'\over 4\pi}\ln~[{(x_1-x_2)^2+a(t)^2\over (x_1-x_2)^2+a(0)^2}]
\end{equation}
is like a propagator. Note $F(x_1,x_2,t)$ will also be denoted by $F_{x_1x_2t}$ which are distinct from F, which is the coupling of the $\cos b_1.X(x)$ term.
Also
\begin{equation}
\label{fuut}
F(x,x,t)=-{\alpha'\over 4\pi}\ln~{a(t)^2\over a(0)^2}=-{\alpha'\over 2\pi}t
\end{equation}

Thus the evolution operator acting on the $e^{-S_{int}}$ gives 

\be
\label{cumulant}
\psi(t)=exp[\sum_{n=0}^\infty {(-1)^n\over n!}<S^n>_c]
\ee
where $\langle S^n\rangle_c$ stands for the connected part of $<S^n>$ and $<...>$ stands for doing the $X''$ integral. This is the cumulant expansion for the Wilson action at scale t.

\subsubsection{Leading Order $\beta$ function for $F(t)$.}

Let us bring down one power of $(S[X,0])_c=(\int L)_c $, where the sub-script c signifies that only the connected parts for all terms will be retained from $e^{-S[X,0]}$ and act on it with the evolution operator. Writing the cosine as a sum of exponentials, and noting  that the action of the evolution operator gives the same factor for both exponentials, we get: 
\[
(\int L)_c \equiv \int {d^2x_1\over a(0)^2} {F \over 4\pi} e^{\hf (b)^2(F(x_1,x_1,t))}cos ({b X(x_1)})
\]

Powers of $a(0)$ have been added for dimensional consistency. We can use the form given in \eqref{fuut} to get
\[
(\int L)_c \equiv \int {d^2x_1\over a(0)^2}  {F \over 4\pi} ({a(0)^2\over a(t)^2})^{{b^2\over 4}}cos({bX(x_1)})
\]
 The factor $({a(0)^2\over a(t)^2})^{{b^2\over 4}}$ is the effect of self contractions in a range of energies $(\Lambda  , \Lambda e^{-t})$. This 
is also the normal ordering factor that one usually obtains which has $a(t)$ replaced by the IR cutoff $1/m$. The usual normal ordering integrates out self contractions of {\em all} fields, i.e up to the IR cutoff. In the ERG only some fields are integrated out and  after
the ERG evolution the field $X$ only has lower momentum modes in it, and the pre-factor is the effect of integrating out the rest. One more difference is that normal ordering takes care of only self interactions. The ERG removes all interactions between high momentum modes because the modes themselves are integrated out. This is the origin of
terms of the form $b^2 F(x_1,x_2,t)$ in the exponent(which will be seen in the later calculations). This is like the correlator between two exponentials, but with only some modes - high momentum - participating.

This can be written as
\[
\int {d^2x_1\over a(t)^2}   {F \over 4\pi}  \left(  {a(0)^2\over a(t)^2} \right)   ^{{b^2\over 4}-1}cos({bX(x_1)})
\]
which shows that it is exactly marginal for $b^2=4$. If ${b^2\over 4}-1<<1$ we can expand 
\[
\left({a(0)^2\over a(t)^2}\right)   ^{{b^2\over 4}-1}\approx 1 - 2t \left({b^2\over 4}-1\right)
\]

Thus $F (t) = F(0) (1- 2t({b^2\over 4}-1))+...$ valid for small $t$. This also gives the leading term in the $\beta$-function:
\be	\label{betaalpha1}
\beta_F=\dot F(t) =-2({b^2\over 4}-1)F _0 = -2({b^2\over 4}-1)F (t)= -2\delta F
\ee
where we have approximated $F(0)$ by $F (t)$ to this order in $t$ and $F$. Here $\delta=b^2/4-1$ is the deviation of the mass dimension of the cosine from marginality as the theory begins to flow. Thus for $({b^2\over 4}-1)>0$ it goes to zero in the infrared and for $({b^2\over 4}-1)<0$ it is a relevant variable that goes to infinity in the IR. This is the lowest order K-T flow.

The third order contribution, ($O(F^3)$), to the $\beta$ function is calculated in Appendix \eqref{thirdorderbetaF}.

\subsubsection{{$\beta_{\delta}$ -- $\beta$ function for b}
}

Since $b$ multiplies X the latter flow is equivalent to field strength renormalization. So we would like to get terms on the RHS of the ERG involving $cos (bX)$ or $\p X \p X$. One has to bring down the term

\[
{1\over 2!} (\int L )_c^2 
\]

Thus we need to evaluate the action of the ERG operator on

\begin{equation}
( cos(bX(x_1)) cos (bX(x_2)) )_c = {1\over 4}(e^{i b X(x_1) }+ e^{-i b X(x_1)})(e^{i b X(x_2)}+e^{-i b X(x_2)})_c
\end{equation}

It is clear that the product can only give terms whose leading term is $1$ or $e^{2ibX}$. The anomalous dimension of $e^{2ibX}$ is $4b^2/2=2b^2$. For $\cos bX$ to be marginal, $b^2$ has to be set to 4. This gives $2b^2 \approx 8$. For the operator $\cos 2bX$, the deviation from marginality is given by $2b^2-2\approx 6$.(The marginality condition for $\cos bX$ is $b^2/2-2\approx0$). Thus it is a highly irrelevant operator. The term starting with $1$ can have terms involving the marginal $\int d^2 x~\p X \p X$. This corrects the kinetic term which gives the flow for the b parameter in terms of $\delta$.

The action of ERG evolution operator on the marginal combination gives
\[
{1\over 2!}{F^2\over 4(4\pi)^2}\int {d^2x_1\over a(0)^2}\int {d^2x_2\over a(0)^2}~e^{b^2 F(x_1,x_2) +{b^2\over 2} (F(x_1,x_1)+F(x_2,x_2))}(e^{i b X(x_1) -ibX(x_2) }+e^{-i b X(x_1) +ibX(x_2) })
\]

where only the contributing terms have been retained.

Replacing $x_2-x_1=y$ we get
\[
-{b^2\over 16}{F^2\over (4\pi)^2}({a(0)^2\over a(t)^2})^{{b^2\over 2}-2} \int {d^2x_1\over a(t)^2}\int {d^2y\over a(t)^2}({y^2+a(t)^2\over y^2+a(0)^2})^{b^2\over 2}y^2(\p X)^2
\]
We are interested in the logarithmically divergent part in order to match with the continuum calculation. (In the above equation one can also replace $a(0)$ by $a(t_0)$ and pick terms proportional to $ln ~({a(t)\over a(t_0)})$.)  We also take the limit $a(0)\to 0 $ so that all powers of $a(0)$ can be set to zero.  But in the limit $a(0)\to 0$ there is translation invariance in time ($t=\ln ({a(t)\over a(0)})$)in the evolution equation and as explained in the introduction the beta function cares only about the linear term in $t$. Furthermore if we assume that $a(t)\approx {1\over m}$, which is the IR cutoff, we can replace $y^2+a(t)^2$ by $a(t)^2$. Thus we get for the $y$ integral:
\[
\pi (a(t)^2)^{{b^2\over 2}-2} \int d(y^2)({1\over y^2+a(0)^2})^{b^2\over 2}y^2\]

Putting back the prefactors:

\be	\label{z(t)}
=-({\delta+1\over 16}){F^2\over 4\pi}[{({a(t)^2\over a(0)^2})^{-2\delta}- 1\over -2\delta}-{({a(t)^2\over a(0)^2})^{-2\delta -1}-1\over-2\delta -1}]\int d^2x_1 ~\p_aX \p^a X(x_1)
\ee
Let us take the limit $\delta \to 0$ and keep leading terms:
\[
=
-({\delta+1\over 16}){F^2\over 4\pi}[2t +O( t^2\delta)  +( ({a(0)^2\over a(t)^2})(1- 4t\delta +...)-1)(1-2\delta+...)]\int d^2x_1 ~\p_aX \p^a X(x_1)
\]
If we now take $a(0)\to 0$ we get only the first term. The beta function only cares about the leading logarithm, which is the linear term in $t$.
This is a correction to the kinetic term ${1\over 4\pi}\int d^2x_1 ~\p_aX \p^a X(x_1)$. Therefore the beta function is
\[
 \beta _{\delta} = -({\delta+1\over 8})F^2
\]

\subsection{The Beta functions.}

Collecting all the beta functions we get
\br
\beta_F &=& - 2F \delta-{F^3\over 8}\\
 \beta _{\delta}&=&-{F^2\over 8}(1+\delta)
\er

The $O(F^3)$ piece for $\beta_F$ is calculated in Appendix \eqref{thirdorderbetaF}.

\subsection{Comparing with Amit {\it et al} \cite{Amit}.}

In their notation ${\beta^2\over 8\pi} = {b^2\over 4}$. Thus
${\beta^2\over 8\pi}=\delta +1$. This is the same $\delta$ that they use. ${F_A\over \beta^2} ={F\over 4\pi}$ where $F _A$ is the variable used in \cite{Amit}. Thus we have
\[
F= {F_A\over 2(\delta +1)}
\]

If we write $F = {F_A\over 2}$, (which is not quite the same as ${F_A\over 2(1+\delta)}$) we get the beta functions in their notation

\br
\beta_{F_A} &=& - 2F_A \delta-{F_A^3\over 32}\\
 \beta _{\delta_A}&=&-{F_A^2\over 32}(1+\delta)
\er

to first order in $\delta$. We can compare this with the beta functions obtained by Amit {\it et al}.

\br
\beta_F &=&   2F_A \delta+{5F_A^3\over 64}\\
 \beta _{\delta}&=&{F_A^2\over 32}(1-2\delta)
\er

(Their beta functions are given by the flow to the UV
and have the opposite sign.)

The zero-eth order terms agree with \cite{Amit}. The first order terms are not universal.  It is shown in their paper that $B+2A$ is a universal quantity where $A$ and $B$ are the non-leading coefficients. $B+2A={5\over 32}-{2\over 32}={3\over 32}$. It can be checked that we get the same (${2+1\over 32}$).

\section{The Central Charge}
We can use the ERG to compute the central charge of a theory  using the method described in \cite{Friedan,Komargodski}. For completeness we review the basic ideas. Later the same ideas will be used for the sine-Gordon theory.
\subsection{Discussion of Central Charge Calculation}

Let $\hat g_{\alpha \beta}= e^{2\sigma}\delta_{\alpha\beta}$.
As is well known \footnote{See for instance \cite{polyakov,Polch}.}
\be
\int {\cal D}_{\hat g}Xe^{-\hf\int d^2x~(\p X)^2} = e^{{1\over 24\pi}\int d^2x (\p\sigma)^2}
\ee

{\bf Proof:}

We start with the action

\begin{equation}
S[g,X]=\frac{1}{4\pi \alpha'} \int d^2 x      \sqrt{g}    \left( g^{\alpha \beta}     \p _{\alpha} X_{\mu}     \p_{\beta}      X^{\mu}     \right)
\end{equation}

We will analyze how the partition function changes under Weyl rescalings. consider two metric related by the transformation

\begin{equation}
\hat{g}_{\alpha \beta}= e^{2\sigma} g_{\alpha \beta}
\end{equation}

On varying $\sigma$ the partition function $Z[\hat{g}]$ changes as

\begin{equation}
\frac{1}{Z[X,\hat{g}]}\frac{\p Z[X,\hat{g}]}{\p \sigma}=\frac{1}{Z[X,\hat{g}]} \int    D_{\hat{g}} X e^{-S[X,\hat{g}]}           \left(   -  \frac{\p S[X,\hat{g}]}{\p \hat{g}_{\alpha \beta}}   \frac{\p \hat{g}_{\alpha   \beta}}{\p \sigma}              \right)
\end{equation}

\begin{equation}
=\frac{1}{Z[X,\hat{g}]} \int    D_{\hat{g}} X e^{-S[X,\hat{g}]}           \left(   -     \frac{1}{2\pi}           \sqrt{\hat{g}}       T^{\alpha} _{\alpha}        \right)
\end{equation}

Since

\begin{equation}
T^{\alpha}_{\alpha}=-\frac{c}{12} R
\end{equation}

\begin{equation}
\frac{1}{Z}\frac{\p Z}{\p \sigma}=\frac{c}{24\pi}         \sqrt{\hat{g}}         \hat{R}
\end{equation}

For two metrics related by a Weyl transformation $\hat{g}_{\alpha \beta}= e^{2\sigma} g_{\alpha \beta}$, their Ricci scalars are related by

\begin{equation}
\sqrt{\hat{g}} \hat{R}    =        \sqrt{g}      \left(      R     -2   \nabla^2 \sigma                                  \right)
\end{equation}

Therefore,

\begin{equation}
\frac{1}{Z}\frac{\p Z}{\p \sigma}=\frac{c}{24\pi}         \sqrt{g}      \left(      R     -2   \nabla^2 \sigma                                  \right)
\end{equation}

This is a differential equation that expresses the partition function, $Z[\hat{g}]$, defined on one worldsheet, in terms of $Z[g]$, defined on another. Solving this we get

\begin{equation}
Z[\hat{g}]=Z[g]    \exp \left[-\frac{1}{4\pi \alpha'}       \int   d^2 x   \sqrt{g}    \left(  -    \frac{c \alpha'}{6}        (g_{\alpha \beta}     \p^{\alpha}   \sigma    \p^{\beta}      \sigma          +R   \sigma)                       \right)               \right]
\end{equation}

If we set $g_{\alpha   \beta}=\delta_{\alpha    \beta}$ and $\alpha'=\frac{1}{2\pi}$, to match with the kinetic term on the LHS of the statement(where $\sqrt{\hat{g}}$ has been suppressed throughout), then the above equation becomes

\begin{equation}
Z[\hat{g}]=  \exp \left[-\frac{c}{24 \pi}       \int   d^2 x     \left(     \p   \sigma                        \right)   ^2               \right]
\end{equation}

where $\sqrt{g}=1$, $c=1$ for a single scalar and $R=0$.

QED.

Even though $g$ drops out of kinetic term, the information about $\hat g$ comes from defining the operator:
\[
\Delta = {1\over \sqrt g}\p_\alpha \sqrt g g^{\alpha \beta}\p_\beta = e^{-2\sigma}\Box
\]
And what we are calculating is $Det^{-\hf} \Delta$.
It is thus there in the measure.

It is implicit in the above that the UV cutoff is taken to infinity. Thus we can write
\be   \label{M1}
\int  _{\Lambda \to \infty}{\cal D}_{\hat g}Xe^{-\hf\int d^2x~(\p X)^2} = e^{{1\over 24\pi}\int d^2x (\p\sigma)^2}
\ee

On the other hand because of scale invariance, we do not have to take $\Lambda \to \infty$. We can also write
\be	\label{M2}
\int  _{\Lambda \to 0}{\cal D}_{\hat g}Xe^{-\hf\int d^2x~(\p X)^2} = e^{{1\over 24\pi}\int d^2x (\p\sigma)^2}
\ee
without modifying the action, i.e. it is not the Wilson action obtained by integrating out modes from \eqref{M1}.

In flat space we can set $\sigma=0$ in the above to get:
\be   \label{M3}
\int  _{\Lambda \to \infty}{\cal D}Xe^{-\hf\int d^2x~(\p X)^2} = 1=
\int  _{\Lambda \to 0}{\cal D}Xe^{-\hf\int d^2x~(\p X)^2} 
\ee

Thus we can say that for $\Lambda\to \infty$, the following statement about integration measures is true:
\be	\label{M4}
{\cal D}_{\hat g}X~e^{-\hf\int d^2x~(\p X)^2}={\cal D}X~ e^{-\hf\int d^2x~(\p X)^2}~ e^{{1\over 24\pi}\int d^2x (\p\sigma)^2}
\ee

We cannot take finite values of $\Lambda$ because we may have to integrate over expressions that contain a scale.

 Now consider adding a mass term : $\hf\int d^2x~ \sqrt {\hat g} m^2 X^2= \hf\int d^2x~ e^{2\sigma} m^2 X^2$. This term explicitly violates scale invariance. We can add a dilaton to make it Weyl invariant:
$\hf\int d^2x~ e^{2\sigma+2\phi} m^2 X^2 $.
So if we set $\delta \phi = - \delta \sigma$, it is invariant. Thus the invariance is spontaneously broken rather than explicitly. Because of this if we now integrate over $X$ we expect the anomaly to remain the same. Thus we expect

\be   \label{M5}
\int  _{\Lambda \to \infty}{\cal D}_{\hat g}Xe^{-\hf\int d^2x~(\p X)^2 + m^2 e^{2\sigma+2\phi} X^2} = e^{{1\over 24\pi}\int d^2x \hat R \phi - (\p\phi)^2}=e^{{1\over 24\pi}\int d^2x~ 2 \phi \Box \sigma  - (\p\phi)^2}
\ee

Therefore on setting the variation $\delta \phi = -\delta \sigma$ we get $-\delta \sigma{1\over 12\pi}\Box \sigma=\delta \left({(\p \sigma)^2 \over 24\pi}\right)$. Thus we have obtained the original anomaly.

For $\Lambda<<m$,
\be	\label{M6}
\int  _{\Lambda <<m}{\cal D}_{\hat g}Xe^{-\hf\int d^2x~(\p X)^2 + m^2 e^{2\sigma+2\phi} X^2}=1
\ee
because all the modes are frozen - effectively there is no scalar field.

Both equations in flat space ($\sigma=0$) give:
\be   \label{M7}
\int  _{\Lambda \to \infty}{\cal D}Xe^{-\hf\int d^2x~(\p X)^2 + m^2 e^{2\phi} X^2} = e^{-{1\over 24\pi}\int d^2x  (\p\phi)^2}\ee
and
\be	\label{M8}
\int  _{\Lambda <<m}{\cal D}Xe^{-\hf\int d^2x~(\p X)^2 + m^2 e^{2\phi} X^2}=1
\ee

Here, the coefficient of the dilaton kinetic term $(\p \phi)^2    \over 24\pi$ in $ e^{-{1\over 24\pi}\int d^2x  (\p\phi)^2}$ in \eqref{M7} is the anomaly of the defining UV theory because of the Weyl violating mass term $m^2X^2$.
Under an RG flow from $\Lambda =\infty$ to $\Lambda=0$ we should get the anomaly, that we get for the defining theory in the UV, from the Wilsonian action in the IR. Thus we should get 
\be	\label{M9}
\int  _{\Lambda <<m}{\cal D}Xe^{-\hf\int d^2x~(\p X)^2 + m^2 e^{2\phi} X^2 + \Delta L[\phi]}=e^{-{1\over 24\pi}\int d^2x~    (\p\phi)^2}
\ee

Here $\Delta L(\phi)$ are all the additional terms in the Wilson action that are generated under an RG flow to the IR. But since effectively there is no integration and all degrees are frozen, we must have 
\[
\Delta L =-{1\over 24\pi}~    (\p\phi)^2
\]
This gives the expected result $\Delta c=1$ for a free massive scalar as you flow from the UV to the IR.

\subsection{  Central Charge for Free Scalar: }

 Let us now apply the ERG evolution operator to obtain the $\phi$ dependence along the RG trajectory. This gives us a definition of the $c$-function.

 We start with a non-local action
\be
S = -\int d^2u \int d^2v ~\hf z(u,v,t) X(u) X(v) - m_0(t)
\ee

But then we choose $z(u,v,0)= \delta m^2 (u) \delta(u-v)$ as our bare action at $t=0$ and then set $\delta m^2(u) = (e^{2\phi(u)}-1)m^2$ where $\phi$ is the external dilaton field.

\[
S[\phi] = \int d^2x~ \hf m^2 X^2 (e^{2\phi}-1)
\]
We act with the evolution operator on the interaction term.
\[
\int {\cal D} X' e^{-\hf\int d^2x_1 \int d^2x_2F^{-1}(x_1,x_2)X'(x_1)X'(x_2)}e^{-S[X+X']}
\]

This is the integral form of the evolution operator obtained in \eqref{ergintegral form}.
We are interested in the coefficient of $(e^{2\phi}-1)^2$ because one has to extract the coefficient of the dilaton kinetic term which gives the c function and this is the term which will contribute to the leading order. We set $X=0$ and evaluate
\[
\int {\cal D} X' e^{-\hf\int d^2x_1 \int d^2x_2[F^{-1}(x_1,x_2)X'(x_1)X'(x_2) +  m^2 X'^2 (e^{2\phi}-1)]}
\]
Path integrating we get
\[
e^{-\hf Tr Ln [F^{-1}+ m^2 (e^{2\phi}-1) ]}=e^{-\hf (Tr Ln [F^{-1}]+ Tr Ln[1+ F m^2 (e^{2\phi}-1) ])}
\]Expanding the logarithm one gets for the quadratic (in $\phi$ ) term:
\[
{1\over 4}Tr [(F m^2 (e^{2\phi}-1))^2]\]
\be	\label{cfn}
={1\over 4} m^4\int d^2x_1 \int d^2x_2 \phi(x_1)(x_1-x_2)^2\p^2 \phi(x_1)(G(x_1,x_2,a(t))-G(x_1,x_2,a(0)))^2
\ee

$G(x_1,x_2,a(t))$ is understood to be evaluated with a cutoff equal to $a(t)=a(0)e^{t}$. $t\to \infty$ corresponds to $a(t) =\infty$. All modes have been integrated out. So the propagator vanishes: $G(x_1,x_2,\infty)=0$. This is also clear from \eqref{RegGF}. When $t=0$ we get the propagator at the UV scale $a(0)$. So we go from the completely unintegrated theory at $a(t=0)$ to one with everything integrated out at $a(t\to \infty)$. Integrating by parts we get,
 \be	\label{C}
 -{1\over 4} \int d^2x_1 \int d^2x_2 [G(x_1,x_2,0)]^2m^4 (x_1-x_2)^2(\p \phi)^2
 \ee

Now
\[
G(x_1,x_2,0)= {1\over 2\pi} K_0(m \left\lvert x_1-x_2  \right\rvert )
\] 

Substituting in \eqref{C}
we get
\be	\label{C1}
-{1\over 24\pi}\int d^2x~(\p \phi)^2
\ee

What we have calculated is  $-L(u,\infty)+L(u,0)= -\Delta L=1$. The change in $c$ is thus 1. The final theory where the scalar field is infinitely massive has $c=0$. The initial theory therefore had $c=1$. The anomalous transformation under scale changes is provided by the
$(\p\phi)^2$ term - this is the argument used by \cite{Friedan,Komargodski}. We have obtained it using the ERG.

\eqref{cfn} defines a c-function for any value of $t$ along the flow. It is also clear that it is monotonic.

\subsection{Central Charge of the Sine-Gordon Theory}

In this section we calculate yet another flow - the c-function defined by Zamolodchikov. We calculate it using the ERG first. We also compare this with a calculation using a prescription given in \cite{Komargodski}. 

\subsubsection{The C-function using the ERG flow equation}

Let us begin with 
some normalization details. The interaction vertex is,

\[
\int d^2x  {F\over a(0)^2} cos~b X
\]then. This term violates Weyl invariance and therefore one introduces a dilaton to restore Weyl invariance.

\[
\int d^2x   e^{2\phi} {F\over a(0)^2} cos~b X
\]

Therefore, under $\sigma \to \sigma + \xi$ and $\phi \to \phi -\xi$ the theory is invariant. $ a(0) \to a(0)e^{-\phi}$ gives the dilaton coupling. Instead of associating the dilaton with $a(0)$ we associate it to the coupling constant $F$ or equivalently to the dimensionful operator $cos~b X$. When we do RG evolution $a(0)\to a(0)e^t \equiv a(t)$ and as usual and there is no $\phi$ associated with it.

{\bf  The dilaton coupling:}

The normal ordered interaction term is

\[
S= \int {d^2x\over a(0)^2} F ~\left({a(0)\over a(t)}\right)^{\frac{b^2}{2}} : cos~ b X:
\]

\[
= \int {d^2x\over a(t)^2} F ~\left({a(0)^2\over a(t)^2}\right)^\delta : cos~ b X:
\]

and $\delta = {b^2 \over 4}-1$ as before. Now we introduce a $\phi$  dependence, we get
\[
S=\int {d^2x\over a(t)^2} F ~e^{-2\delta (t+\phi(x,t))} cos~ b X
\]

To this order

\[
F(t,\phi)=F (\phi) e^{-2\delta t}=F e^{-2\delta (t+\phi(x))}
\]
Note that the coupling constant has become $x$-dependent and has to be placed inside the integral:
\[
\int {d^2 x\over a(t)^2}~ F (t,x)~cos~b X(x)
\]

Thus we have determined the dilaton coupling. This has the information of the contribution of the anomalous scaling behaviour of the cosine operator under an RG flow to the central charge.

{\bf Extracting the anomaly:}

As discussed before, the anomaly is the coefficient of the dilaton kinetic term. One has to go to over to the ${1\over 2!}<S^2>_c$ term, where the subscript c signifies taking only the connected parts, to extract the anomaly. We act on this term by the ERG operator and extract the dilaton kinetic term. The calculation proceeds as follows,

\[
{1\over 2!} \int {\cal D} X''e^{\hf \int d^2x_1~\int d^2x_2~F^{-1}(x_1,x_2,t)X''(x_1)X''(x_2)}\]
\[F^2 \int _x  [{e^{ib X(x)+ib X''(x)}+e^{-(ib X(x)+ib X''(x))}\over 2}]\int _y [{e^{ib X(y)+ib X''(y)}+e^{-(ib X(y)+ib X''(y))}\over 2}]
\]

Here $\int_x=\int \frac{d^2x}{a(0)^2}$.
We replace $F$ by $F (t,\phi)$ as before. a(t) is the IR scale for the action. The propagator has an exponential fall off beyond the IR scale. So when $|y-x|\approx a(t)$ the propagator is highly damped. So we are justified in assuming that $a<|x-y|<a(t)$. Thus the total contribution is (letting $z=y-x$)
\[= 2
{F^2\over 8}\int {d^2x\over a(t)^2}~\int {d^2z\over a(t)^2}e^{-2\delta[2t+\phi(x,t)+\phi(y,t)]}e^{{b^2 \over 2}ln~ \left( {a(t)^2\over z^2+a(0)^2}\right)}
(1+ izb \p X...)\]

Now,

\[
e^{-2\delta(\phi(x)+\phi(y))}=1 - 2\delta(\phi(x)+\phi(y)) +2\delta^2(\phi(x)+\phi(y))^2+...
\]
The relevant part is 
\be	\label{phi}
2\delta^2 2 \phi(x) \phi(y) =2\delta^2\phi(x) (y-x)^a (y-x)^b \p_a \p_b \phi(x)
\ee
 Inserting \eqref{phi}  for $e^{-2\delta(\phi(x)+\phi(y))}$ we get for the term in the Wilson action involving $\phi \Box \phi$:

\be	\label{total}
=\int d^2x~{F^2(t)\over 4}\delta^2\phi(x) \p^2\phi(x)(a(t)^2)^{2\delta}\int d^2z ~z^2(z^2+a(0)^2)^{-b^2 \over 2}
\ee

Here we have used rotational symmetry to replace $z^a z^b $ by $z^2{\delta^{ab}\over 2}$. The integral is log divergent and the divergent piece can be extracted by introducing the regulator $a(0)$ in the limits rather than in the integrand: (The IR end is cutoff anyway by $a(t)$.)
\[
\int d^2z ~z^2(z^2+a(0)^2)^{-b^2 \over 2}=
\pi {[(a(0)^2)^{-2\delta} - (a(t)^2)^{-2\delta}]\over -2\delta}
\]

Inserting in \eqref{total} and expanding for small $\delta$ we get

\be    \label{97}
-\pi\int d^2x~{F^2\over 2}\delta^2\phi(x) \p^2\phi(x) t
\ee
The answer depends on the logarithmic range $t$. The calculation can be improved if we realise that $F $ is a function of $t$. We assume that the range of RG evolution $t$ is infinitesimal - $dt$. Then we can replace $t \to \int_0^t dt$ and acknowledge the functional dependence of F on t explicitly, $F(t)$. Then we can write \eqref{97} as
\[
-\pi\int d^2x~\delta^2\phi(x) \p^2\phi(x) \int dt {F^2(t)\over 2} 
\]
Noting that ${dF\over dt}=-2\delta F$ we can write $dt =- {dF \over 2\delta F}$ to get

\[=
-3\pi^2F^2 \delta\times{1\over 24\pi}\int d^2x~\p_a\phi(x) \p^a\phi(x)
\]
The coefficient of $-{1\over 24\pi}\int d^2x~\p_a\phi(x) \p^a\phi(x)$ gives the change in the central charge.
Thus
\[
\Delta c= c(F(0))-c(F(t)) = 3\pi^2F^2 \delta
\]

Here $c(F(0))$ if the central charge of the UV theory. $c(F(t))$ is the central charge of the IR theory. When $\delta >0$ we have an irrelevant operator - $F$ flows to zero under an RG evolution. So $c(F(0))>c(F(t))$ - which is correct.

\subsubsection{A confirmation with a result from Entanglement Entropy}

At the conformal point the entanglement entropy for a single interval is related to the central charge of the CFT\cite{Calabrese:2004eu} by

\begin{equation}
 EE ={ c\over 3}\ln (l/\epsilon) 
\end{equation}

where $l$ is the length of the interval and $\epsilon$ is the short distance cutoff. If you identify a(t) with $l$ and a(0) with $\epsilon$ and then analyze the behaviour of this expression as an RG flow, then to leading order, one would expect the change in entanglement entropy when one goes slightly away from the fixed point to be

\begin{equation}
\Delta EE ={\Delta c\over 3}\ln \left(\frac{a(t)}{a(0)}\right) +H.O.T.
\end{equation}

Substituting the $\Delta c$ above we get,

\begin{equation}
\Delta EE = \pi^2 F^2 \delta   \ln \left(\frac{a(t)}{a(0)}\right)
\end{equation}

If we set $F = {\lambda\over 8\pi}$ and $2\delta = \Delta-2$ we get
\[
\Delta EE =  {\lambda^2\over 128}( \Delta-2)~ \ln \left(\frac{a(t)}{a(0)}\right)
\]

$\Delta EE$ has recently been calculated holographically in \cite{BBS}. We show that this expression is in agreement with their results.

\subsubsection{The c-function from Komargodski's prescription}

Now we will calculate the c-function of the Sine-Gordon using a technique by Komargodski and show that the results match with our earlier calculation. This is a check on both techniques.
The interaction term for the Sine-Gordon action is

\begin{equation}
S=\int {d^{2} x  \over a(0)^{2}} F  cos( b X(x)) 
\end{equation}

\begin{equation}
S=\int  {d^{2} x  \over a(0)^{2}} F  :cos( b X(x)): (ma(0))^{(b^2/2)}
\end{equation}

Here the scale dependence to the lowest order from normal ordering has been explicitly factored out. $m$ will be identified with $\frac{1}{a(t)}$, where $a(t)$ is the UV cutoff after several RG transformations have been performed and is thus the IR scale.

\begin{equation}
S=\int m^{2} d^{2} x F  :cos( b X(x)): (ma(0))^{(b^2/2-2)}
\end{equation}

If under scaling $a(0)\rightarrow \lambda a(0)$, then under scaling, a dilaton $\exp(\phi)$, would transform as $\exp(\phi)\rightarrow \frac{\exp(\phi)}{\lambda}$, 
thus leaving the action invariant under scaling. We introduce the dilaton in the action and take its effect under scalings into account,

\begin{equation}
S=\int   m^2       d^{2} x F  :cos( b X(x)): (ma(0) \exp(\phi))
^{(b^2/2-2)}
\end{equation}

The Green's function for a massive scalar is
\begin{equation}
G(x_1,x_2)=- \ln    \big(m^2 ((x^2+a(0)^2)\big)
\end{equation}

where $x=x_1-x_2$. The trace of the Energy-Momentum tensor, $T= T^{a} _{a}$ is 

\begin{equation}
T= \frac{\delta S}{\delta \phi}= \int m ^2    d^{2} x F    : cos( b X(x))    :      (ma(0) \exp(\phi))^{(b^2/2-2)} (b^2/2-2)
\end{equation}

\begin{equation}
<T(y) T(0)>= F^{2} \big(b^{2}/2-2   \big)^{2}  (ma(0))^{2(b^2/2)}  <	\cos b X(y) \cos b X(0)> 
\end{equation}

 Komargodski's prescription \cite{Komargodski} for the change in the central charge under an RG flow gives,

\begin{align}
\Delta c &= -3 \pi \int d^{2}y y^{2} < T(y) T(0)> \nonumber \\
&=      \frac{3}{2}   \pi^{2} F^{2} (ma(0))^{b^2-4}  (b^{2}/2-2)         
\end{align}

Substituting $b^{2}/2-2 = 2 \delta$ and identifying $m^{-1} \to a(t)$ ,

\begin{equation}
\Delta c= 3/2 \pi^{2} F^{2} (a(0)/a(t))^{4 \delta} 2 \delta
\end{equation}

Then,
\begin{align}
F^{2} (a(t)/a(0))^{-4 \delta} 2 \delta =F^{2} 2 \delta \bigg (1+ (-4\delta) \ln (a(t)/a(0)) \bigg)
\end{align}

Now we can write $a(t)=a(0) e^t$ where $a(0)$ is the UV cut-off. So we can write $\ln (a(t)/a(0)) $ as $\int_0^t dt$ for t infinitesimal and then promote $F \rightarrow F(t)$.
We get

\begin{align}
\label{cintrmdt}
F^{2} 2 \delta(-4\delta)\ln (a(t)/a(0))  \rightarrow    -8\delta^{2}   \int_0^t          F(t)^{2}     dt	
\end{align}

Therefore we get

\begin{equation}
\label{deltacunint}
\Delta c= -12\pi^2  \delta^2 \int F^2(t) dt
\end{equation}

The Beta function is given by,
\begin{equation}
 \frac{dF}{dt}= \beta_{F}=-2 F \delta
\end{equation}

to leading order. Substituting this in \eqref{cintrmdt} we get

\begin{equation}
\label{eq:Deltac1}
\Delta c           =          3     \pi^{2} F^{2}         \delta
\end{equation}

as before.

\subsection{Higher order terms for $\Delta c$}

Under a change in renormalization $\delta$ goes to(equation 7.6\cite{Amit})

\begin{equation}
\label{eq:delta}
\delta=\delta_0+ a F^2
\end{equation}

Therefore

\begin{equation}
\label{eq:ddelta}
\beta_\delta= \frac{d \delta}{d t}= 2 F a \frac{d F}{dt}= 2 F a \beta_F
\end{equation}

We know

\begin{equation}
\label{eq:betadelta}
\beta_\delta=\frac{d\delta}{dt}=-\frac{F^2}{32}+\frac{F^2\delta}{16}
\end{equation}

and

\begin{equation}
\label{eq:betaalpha}
\beta_F=-2 F \delta- \frac{5}{64} F^3
\end{equation}

where $\beta_F$ and $\beta_\delta$ are the $\beta$ functions as obtained in \cite{Amit}. Substitute \eqref{eq:betaalpha}, \eqref{eq:betadelta} and \eqref{eq:delta} in \eqref{eq:ddelta} we get

\begin{equation}
\frac{F^2}{32}-\frac{F^2 \delta_0}{16}= 2 F a (2 F \delta_0 +\frac{5}{64} F^3)
\end{equation}

Comparing coefficients 

\begin{equation}
a=\frac{-1}{64}
\end{equation}

So

\begin{equation}
\label{eq:deltafinal}
\delta=\delta_0-\frac{1}{64} F^2
\end{equation}

where the $F$ dependence of $\delta$ has been determined to leading order. So,

\begin{align}
& \int \delta^2 F^2 dt =- \int \frac{\delta^2 F dF}{2 \delta+\frac{5}{64}F^2}
\end{align}

where \eqref{eq:betaalpha} has been substituted in the above. Substituting \eqref{eq:deltafinal} in the above expression, simplifying and resubstituting this expression in \eqref{deltacunint} we get

\begin{equation}
\label{eq:Deltac2}
\Delta c=           12 \pi^2      \bigg       (\frac{\delta_0 F^2}{4}- {7F^4 \over 1024} + H.O.T.       \bigg     )
\end{equation}

To lowest order \eqref{eq:Deltac2} matches \eqref{eq:Deltac1}.

\section{Summary and Conclusions}

In this paper we have studied the RG flow of quantities in field theories. The idea is to use Polchinski's ERG written in terms of an evolution operator. The advantage is that one can directly obtain quantities such as the beta function by looking at the linear dependence on the RG time $t$.  In the limit of cutoff going to infinity this coefficient gives the beta function. This technique was illustrated with a few examples such as the $\phi^4$ theory in four dimensions and also the Sine-Gordon theory in two dimensions - which is the main interest in this paper. We also show that another flow calculation that this method is suited for is that of the c-function. We illustrate it with the case of the free field. We then calculate it for the Sine-Gordon theory. Interestingly if we assume the relation between entanglement entropy and the central charge continues to hold even for the c-function we can evaluate the entanglement entropy of the Sine Gordon theory for small values of the perturbation. This has been done using other field theoretic and also holographic methods and there is complete agreement for the lowest order term - which is all that has been calculated\cite{BBS}. For the Sine-Gordon theory the detailed results of \cite{Amit} for the solution of the RG equations has been used in this paper to calculate the c-function to higher orders.

There are many open questions. It would be interesting to extend the ideas in this paper to more basic issues in holographic RG and in particular the connection with the RG on the boundary theory. In the context of entanglement entropy it would be interesting to check the match to higher orders. Since the c-function is presumably not a universal quantity (there should be some scheme dependence) at higher orders,
these checks have to be made keeping these caveats in mind. 

We hope to report on some of these issues soon.

\begin{appendix}

\section{Third Order term in the Sine-Gordon $\beta$-function}\label{thirdorderbetaF}

The third order term is made of two positive exponentials and one negative one or vice versa and there are three such 
terms that can combine to give a cosine as the leading term in the OPE:
\[
3\times{1\over 3!}{F^3\over (4\pi)^3} \int {d^2x_1\over a(0)^2}\int {d^2x_2\over a(0)^2}\int {d^2x_3\over a(0)^2}
{1\over 4}e^{{b^2 \over 2}(F(x_1,x_1)+F(x_2,x_2)+F(x_3,x_3))}
\]
\be	\label{I}
e^{-b^2(F(x_1,x_2)+F(x_1,x_3)-F(x_2,x_3))} cos~(bX(x_1))
\ee
We will set $b^2=4$ without further ado. We choose $x_1=0$ (by translational invariance) and for notational simplicity set $x_2=x,~x_3=y$. As before we choose $a(t)\approx {1\over m}$ so the integral in this approximation becomes (suppressing all prefactors ${1\over 8} {F^3\over (4\pi)^3}$):
\be	\label{II}
\int {d^2x\over a(t)^2} \int {d^2 y\over a(t)^2}~[{a(t)^2\over x^2+a(0)^2}]^2[{a(t)^2\over y^2+a(0)^2}]^2e^{2ln~[{ (x-y)^2+a(0)^2\over a(t)^2}]}
\ee
There are three regions of divergences:
\begin{enumerate}
\item I : $x\to 0,~y > \Delta$

\item II:$y\to 0,~y>\Delta$

\item III: $x,y \to 0$

When both $x,y >\Delta$ and $x\to y$ there is a divergence, but it is of the same form as I or II and is merely a permutation of indices: In the above we have kept $x_1=0$ fixed, but there are other choices which will produce 
three similar regions and this region ($x,y >\Delta$ and $x\to y$) will be one of those.

Here $\Delta$ is some finite arbitrary length. The coefficient of the divergence cannot depend on $\Delta$ because it is an arbitrary way to split up the region of integration.

\end{enumerate}

{\bf Region I:} Let us Taylor expand the log in the last factor, about $y^2$, which is large:
\[
e^{2ln~[{ (x-y)^2+a(0)^2\over a(t)^2}]}=e^{2ln~[{ y^2 +a(0)^2 +\overbrace{x^2 - 2 x.y}^{X}\over a(t)^2 }]}
\]
\[=
e^{2(ln~[{ y^2 +a(0)^2\over a(t)^2}] + {X\over y^2+a(0)^2} -{1\over 2!} X^2{1\over (y^2+a(0)^2)^2}+...)}
\]

Insert this into \eqref{II} and we get:

\[
\int {d^2x\over a(t)^2} \int {d^2 y\over a(t)^2}~[{a(t)^2\over x^2+a(0)^2}]^2\{1+ \underbrace{{2X\over y^2+a(0)^2}}_{(i)} +\underbrace{X^2{1\over (y^2+a(0)^2)^2}}_{(ii)}+..  \}
\]

The leading term in this expansion corresponds to a disconnected graph where $x$ and $0$ are connected and $y$ is not connected to either of these. This has to be subtracted out since, the cumulant expansion prescription is to calculate connected graphs. So we are left with (i) and (ii). The $x^4$ term is finite (on doing the $x$ integral). We get for Region I

\[
-4\pi^2 ln~[m^2\Delta^2]ln~[{\Delta^2\over a(0)^2}]
\]

{\bf Region II:} Gives the same as above.

Thus the total contribution from Region I and II =
\[
-8\pi^2 ln~[m^2\Delta^2]ln~[{\Delta^2\over a(0)^2}]
\]

Notice that there is no $\Delta$ independent contribution to $ln~a(0)$. That can come only when all three vertices
are together. This will come from region III.

{\bf Region III:}

We go back to \eqref{II} in this region of integration. \be	
\int_0^\Delta {d^2x\over a(t)^2} \int _0^\Delta{d^2 y\over a(t)^2}~[{a(t)^2\over x^2+a(0)^2}]^2[{a(t)^2\over y^2+a(0)^2}]^2[{ (x-y)^2+a(0)^2\over a(t)^2}]^2
\ee

We expand
\[
[{1\over x^2+a(0)^2}][{1\over y^2+a(0)^2}][(x-y)^2+a(0)^2]= {1\over x^2+a(0)^2} +{1\over y^2+a(0)^2} - {\overbrace{a(0)^2+2x.y}^{Y}\over (x^2+a(0)^2)(y^2+a(0)^2)}
\]

Squaring it produces six terms:
\begin{enumerate} [label=(\alph*)]

\item 
\[({1\over x^2+a(0)^2})^2\]

\item 
\[({1\over y^2+a(0)^2})^2\]

\item 
\[
{Y^2\over (x^2+a(0)^2)^2(y^2+a(0)^2)^2}
\]

\item 
\[
{2\over (x^2+a(0)^2)(y^2+a(0)^2)}
\]
\item 
\[
{-2Y\over (x^2+a(0)^2)(y^2+a(0)^2)^2}
\]

\item 
\[
{-2Y\over (x^2+a(0)^2)^2(y^2+a(0)^2)}
\]

\end{enumerate}
Terms (a) and (b) correspond to disconnected diagrams that are subtracted out. 

(c)\[
=2\pi^2 [ ln^2~({\Delta ^2\over a(0)^2})+1-2 ln ~({\Delta^2\over a(0)^2})+ O(a^2)]
\]

(d) 
\[
=
2\pi^2 ln^2~{\Delta^2\over a(0)^2}
\]

(e)

\[=-2\pi^2ln~{\Delta^2\over a(0)^2}
\]

(f)=(e)

\[=-2\pi^2ln~{\Delta^2\over a(0)^2}
\]

Any renormalizable theory cannot have divergences of the type $ln~ \Delta^2 ln ~a(0)^2$. Because $\Delta$ is like momentum and the counter terms would have derivative interactions to all orders. Thus the theory would be non-local.

We can now check that the coefficient of $ln~ \Delta^2 ln ~a(0)^2$  is zero.

\[ 
8\pi^2 ~(from~ I+II)~ - 4\pi^2 ~(from ~(c)) ~- 4\pi^2 ~(from ~(d)~ =0
\]

The coefficient of $ln~a(0)^2$ is $8\pi^2$.
Thus we get putting back the prefactors
\[
{1\over 8}{F^3\over (4\pi)^3} 8\pi^2 2~ln~{a(0)\over \Delta}
\]
Since any value of $\Delta$ is safe for extracting the divergence, we can extend the region of
integration to its full value which is $\Delta= a(t) \approx {1\over m}$.

This the modified $F\over 4\pi$. So
\[
 F (t)={F^3\over 8}~ln~{a(0)\over a(t)} 
\]

Thus the beta functions at this order is

\be	\label{betaalpha2}
\beta_{F}= -{F^3\over 8}
\ee

\end{appendix}

\end{document}